\documentclass[12pt]{iopart}

\begin{document}

\title[Letter to the Editor]{Melosh rotation: source of the proton's missing spin}

\author{Bo-Qiang Ma}

\address{Center of Theoretical Physics, CCAST (World Laboratory), Beijing, People's Republic of China
\\and \\Institute of High Energy Physics, Academia Sinica, PO Box 918
(4), Beijing 100039, People's Republic of China}

\begin{abstract}
It is shown that the observed small value of the integrated spin
structure function for protons could be naturally understood within
the naive quark model by considering the effect from Melosh
rotation. The key to this problem lies in the fact that the deep
inelastic process probes the light-cone quarks rather than the
instant-form quarks, and that the spin of the proton is the sum of
the Melosh rotated light-cone spin of the individual quarks rather
than simply the sum of the light-cone spin of the quarks directly.

\end{abstract}


\section*{}
The spin content of the proton has received extensive attention from
the particle physics community recently. The reason for this is that
the European Muon Collaboration (EMC) found~\cite{Ashman} from their
polarized muon proton data a much smaller value of the integrated
spin structure function for protons compared with that from the
Ellis-Jaffe sum rules~\cite{Ellis}. This small integrated spin
structure function combined with the Bjorken sum rule~\cite{Bjorken}
was interpreted as the evidence that a very small fraction of the
proton's spin is provided by the spin of its quarks. This
conclusion, if true, is of course startling because it is in clear
contradiction with the previous theoretical expectations
~\cite{Jaffe}. Hence, many papers have been devoted to this problem
and many complicated models for the proton's missing spin have been
proposed. In this letter, we indicate, however, that the small value
of the proton's integrated spin structure function could be
naturally understood in the naive quark model ({\tiny NQM}) by
considering the effect of the Melosh
rotation~\cite{Melosh,Berestetskii}.

The key to this problem lies in two very simple but in practice
often mistakenly treated or ignored facts. The first is that the
deep inelastic lepton scattering process is a probe of the
light-cone (or current) quarks rather than the instant-form (or
constituent) quarks~\cite{Drell,Brodsky,Ma}. The second is that in
light-front dynamics the spin of the proton is not simply the sum of
the spin of the individual quarks but the sum of the Melosh rotated
spin of the light-cone quarks~\cite{Coester,Chung}. The theoretical
bases for the two facts can be traced back to the old work of
Dirac's relativistic Hamiltonian dynamics~\cite{Dirac}, Weinberg's
infinite momentum technique~\cite{Weinberg}, and Wigner's spin state
rotation~\cite{Wigner}. The first fact leads naturally to the
conclusion that the quark's spin measured in deep inelastic lepton
scattering is the light-cone spin rather than the instant-form spin.
Taken in conjunction with the second fact, we can conclude that
there is no need to require that the sum of the quark's spin
measured in deep inelastic process be equal to the proton's spin.

In the following, we simply present an intuitive model to evaluate
the effect from Melosh rotation. We start from the conventional
instant-form (T) {\tiny NQM} SU(6) proton wavefunction
\begin{equation}\label{eq1}
|p^\uparrow_T\rangle=(2u^\uparrow_Tu^\uparrow_Td^\downarrow_T-u^\uparrow_Tu^\downarrow_Td^\uparrow_T-u^\downarrow_Tu^\uparrow_Td^\uparrow_T)/\sqrt{6}
~(+\mathrm{cyclic ~permutation})
\end{equation}
one finds $\triangle u_T=\frac{4}{3}$, $\triangle d_T=-\frac{1}{3}$
and $\triangle s_T=0$. If the deep inelastic process is a probe of
the instant-form quarks, we expect, respectively, the integrated
spin structure function for protons
\begin{equation}\label{eq2}
\int \textrm{g}_1^pdx=\frac{1}{2}(\frac{4}{9}\triangle
u+\frac{1}{9}\triangle d)=\frac{5}{18}=0.278
\end{equation}
and that for neutrons
\begin{equation}\label{eq3}
\int \textrm{g}_1^ndx=\frac{1}{2}(\frac{1}{9}\triangle
u+\frac{4}{9}\triangle d)=0
\end{equation}
together with the proton's spin sum rule
\begin{equation}\label{eq4}
(\triangle S^T_Z)_{u+d+s}=\frac{1}{2}(\triangle u_T+\triangle
d_T)=\frac{1}{2}
\end{equation}
which means that the proton's full spin is carried by its valence
quarks.

 The instant-form (T) quark states $q^s_T$ and the front-form (F) quark states $q^s_F$
 are related by the Melosh rotation~\cite{Melosh,Berestetskii,Coester,Chung}
\begin{equation}\label{eq5}
q^s_F=\sum_{s'} M_{s's}(R)q^{s'}_T
\end{equation}
with the Melosh rotation operator defined by
\begin{equation}\label{eq6}
R=(m+k_0+k_3+i\varepsilon_{ij3}\sigma_ik_j)/[2(k_0+k_3)(m+k_0)]^{1/2}
\end{equation}
in specifying $q^s_F$ and $q^s_T$ by the two-component Pauli
spinors. From (\ref{eq5}), we get, inversely,

\begin{equation}
q^\uparrow_T =w[(k^++m)q^\uparrow_F-k^Rq^\downarrow_F] \hspace{50pt}
q^\downarrow_T =w[(k^++m)q^\downarrow_F+k^Lq^\uparrow_F]\label{eq7}
\end{equation}
in which $w=[2k^+(m+k_0)]^{-1/2}$, $k^{R,L}=k_1 \pm ik_2$,
$k^+=k_0+k_3$ and $k_0=(m^2+\textbf{k}^2)^{1/2}$. We see from
(\ref{eq7}) that the light-cone spin carried by an instant-form
quark should be its instant-form spin multiplied by a factor
\begin{equation}\label{eq8}
M_q=[(k^++m)^2-\textbf{k}^2_\bot]/[2k^+(m+k_0)].
\end{equation}
Therefore we can identify
\begin{equation}\label{eq9}
\bigtriangleup q_F=\langle M_q\rangle \bigtriangleup q_T
\end{equation}
in which $M_q$ is the contribution from the Melosh rotation.

We simply assume that the quark momentum-space wavefunction of the
proton is described by the harmonic oscillator wavefunction
\begin{equation}\label{eq10}
\Psi(\textbf{k})=\pi
^{-3/4}\alpha^{-3/2}\exp(-\textbf{k}^2/2\alpha^2).
\end{equation}
We know from previous work~\cite{Isgur} that this wavefunction is
good in describing the static properties of hadrons at low momentum
scale in adopting the harmonic scale $\alpha\approx330$ MeV and the
quark mass $m\approx330$ MeV. At high momentum scale one may expect
that $m$ becomes smaller or $\alpha$ becomes larger~\cite{Isgur}.
Hence the expectation value of $M_q$  may be evaluated by
\begin{equation}\label{eq11}
\langle M_q\rangle=\int d^3\textbf{k}M_q|\Psi(\textbf{k})|^2.
\end{equation}

As deep inelastic scattering is a probe of the light-cone quarks, we
should use $\triangle q_F$ rather than $\triangle q_T$ in equations
(\ref{eq2})-(\ref{eq3}) to calculate the integrated spin structure
functions for protons and neutrons. Assuming $M_u= M_d$ and adopting
$m$, $\alpha$ to be 134 MeV, 330 MeV or 330 MeV, 815 MeV
respectively, we obtain
\begin{equation}\label{eq12}
\int \textrm{g}_1^pdx=\frac{1}{2}(\frac{4}{9}\triangle
u_F+\frac{1}{9}\triangle d_F)=\frac{5}{18}=0.126
\end{equation}
and
\begin{equation}\label{eq13}
\int \textrm{g}_1^ndx=\frac{1}{2}(\frac{1}{9}\triangle
u_F+\frac{4}{9}\triangle d_F)=0
\end{equation}
together with
\begin{equation}\label{eq14}
(\triangle S^F_z)_{u+d}=\frac{1}{2}(\triangle u_F+\triangle
d_F)=0.227
\end{equation}
which means that the sum of the light-cone spin of the valence
quarks is only 45.4\% of the proton's spin. From (\ref{eq12}) we see
that the above intuitive picture could naturally explain the small
EMC data of the proton's integrated spin structure function with
reasonable parameters.

One can easily find that the above results are quantitatively
inconsistent with the Bjorken sum rule. This comes from the adoption
of the {\tiny NQM} SU(6) proton wavefunction and the assumption
$M_u= M_d$. Actually the proton's instant-form wavefunction should
be
\begin{equation}\label{eq15}
|p_T\rangle=a_0|uud\rangle_T+a_1|uudq\bar{q}\rangle_T+a_2|uudg\rangle_T+\cdot\cdot\cdot
\end{equation}
in which the high Fock state contributions could change
$\bigtriangleup u_T$ and $\triangle d_T$ from the values
$\frac{4}{3}$ and $-\frac{1}{3}$. $\langle M_u\rangle$ and $\langle
M_d\rangle$ may also be different since there are two $u$ valence
quarks and one $d$ valence quark in the proton. Bearing the above
consideration in mind, we start from the most recent EMC data
\begin{equation}\label{eq16}
\int \textrm{g}_1^pdx=\frac{1}{2}(\frac{4}{9}\triangle
u_F+\frac{1}{9}\triangle d_F)=\frac{5}{18}=0.126
\end{equation}
and the Bjorken sum rule
\begin{equation}\label{eq17}
\int (\textrm{g}_1^p-\textrm{g}_1^n)dx=\frac{1}{6}(\triangle
u_F-\triangle d_F)=\frac{1}{6}\textrm{g}_A/\textrm{g}_V
\end{equation}
with $\textrm{g}_A/\textrm{g}_V = 1.259$ determined from neutron
$\beta$ decay~\cite{PDG} to evaluate the values of $\triangle u_T$,
$\triangle d_T$, $\langle M_u\rangle$ and $\langle M_d\rangle$. In
order to simplify the discussion, we neglect the possible effects
from the sea or gluon polarization\footnotemark  \footnotetext{Close
~\cite{Close90} indicated recently that the magnitude of the
(strange) sea polarization is likely to be significantly nearer to
zero than is being assumed in much of the current literature. } and
from the quark or gluon orbital angular momentum. From (\ref{eq16})
and (\ref{eq17}), we obtain
\begin{equation}
\triangle u_F=\langle M_u\rangle \triangle u_T=0.705  \hspace{50pt}
\triangle d_F=\langle M_d\rangle \triangle d_T=-0.554\label{eq18}
\end{equation}
and the sum of the light-cone spin of the valence quarks
\begin{equation}\label{eq19}
(\triangle S^F_z)_{u+d}=\frac{1}{2}(\triangle u_F+\triangle
d_F)=0.076
\end{equation}
which is very small. We know that $\triangle u_{T,F}$, $\triangle
d_{T,F}$, $\langle M_u\rangle$ and $\langle M_d\rangle$ should meet
the general requirements
\begin{equation}
-2\leq \triangle u_{T,F} \leq 2 \hspace{20pt} -1\leq \triangle
d_{T,F} \leq 1 \hspace{20pt} 0\leq  \langle M_{u,d}\rangle\leq 1
\label{eq20}
\end{equation}
and the spin sum rule
\begin{equation}\label{eq21}
\frac{1}{2}(\triangle u_T+\triangle d_T)=\frac{1}{2}
\end{equation}
The combination of equations (\ref{eq18}), (\ref{eq20}) and
(\ref{eq21}) leads to the constraints
\begin{eqnarray}
\lefteqn {1.554\leq \triangle u_T \leq 2 \hspace{10pt} 0.352\leq
\langle M_u \rangle\leq 0.454 \hspace{10pt} -1\leq \triangle d_T
\leq -0.554 } \label{eq22}\\
0.554\leq  \langle M_d \rangle\leq 1. \nonumber
\end{eqnarray}

Therefore in order to satisfy both the EMC data and the Bjorken sum
rule simultaneously, it is necessary that $\langle M_u\rangle \neq
\langle M_d\rangle $, $\triangle u_T \neq \frac{4}{3}$, and
$\triangle d_T \neq -\frac{1}{3}$; i.e. the proton's instant-form
valence quark distribution should be different from that of the
{\tiny NQM} SU(6) wavefunction, and the $u$ quark and the $d$ quark
should have different momentum-space wavefunctions in the proton. We
are also interested to see that the Melosh rotation is also an
important source for the depletion of $\textrm{g}_A/\textrm{g}_V$
relative to the value $\frac{5}{3}$ expected from the SU(6) naive
quark model. This is a significant source in comparison with other
sources such as the effect from the quark transverse
momenta~\cite{Bogolaubov} and the effect due to `small' components
in the quark's Dirac spinors in the bag model ~\cite{Close} or in
quark-confining potentials~\cite{Tegen}.

In summary, we present in this letter a very simple model in which
the EMC results of the proton's integrated spin structure function
could be naturally explained within the naive quark model by
considering the effect from Melosh rotation. This model does not
necessarily invalidate the Bjorken sum rule if we impose some
constraints on the Fock state wavefunction of the proton. This work
is based on two very simple hut profound facts which have sound
bases both theoretically and experimentally. Though the quantitative
results in this letter may be changed by the complicated effects
from the sea and gluon polarizations and by contributions from the
orbital angular momentum, or by the anomalous gluon contributions
via the U(l) axial anomaly, the effect from Melosh rotation should
be of fundamental importance in the spin content of hadrons and
therefore should not be ignored. We think the effect revealed in
this letter should have also manifested itself in a number of high
energy processes, and therefore requires further theoretical and
experimental works.


\section*{References}
\begin{thebibliography}{10}

\bibitem{Ashman}
Ashman J {\it et al} (EMC) 1988  {\it \PL} {\bf 206B} 364;
1989 {\it \NP} {\bf B 328 1}

\bibitem{Ellis}
Ellis J and Jaffe R L 1974  {\it Phys. Rev. D} {\bf 9} 1444; 1974
{\it Phys. Rev. D} {\bf 10} 1699 {\bf(E)}

\bibitem{Bjorken}
Bjorken J D 1971  {\it Phys. Rev. D} {\bf 1} 1376;
1966 {\it \PR} {\bf 148} 1467

\bibitem{Jaffe}
Jaffe R L and Manohar A 1990  {\it \NP} {\bf B 337} 509

\bibitem{Melosh}
Melosh H J 1974  {\it Phys. Rev. D} {\bf 9} 1095

\bibitem{Berestetskii}
Berestetskii V R and Terent'ev M V 1976 {\it Yad. Fiz.} {\bf 24}
1044 (Engl. transl. 1976 {\it Sol. J. Nucl. Phys.} {\bf 24} 547) \\
Kondratyuk L A and Terent'ev M V 1980 {\it Yad. Fiz.} {\bf 31} 1087
(Engl. transl. 1980 {\it Sol. J. Nucl. Phys.} {\bf 31} 561)

\bibitem{Drell}
Drell S D, Levy D J, and Yan T -M 1969 {\it \PR} {\bf 187} 2159;
1970 {\it Phys. Rev. D} {\bf 1} 1035\\
Drell S D and Yan T -M 1971 {\it \APNY} {\bf 66} 578


\bibitem{Brodsky}
Brodsky S J 1982 {\it Lectures on Lepton-Nucleon Scattering and
Quantun Chromodynamics} ed A Jaffe and D Rude (Boston:
Birkh\"{a}user) p 255 \\Brodsky S J, Huang T, and Lepage G P, 1983
{\it Particles and Fields} ed A Z Capri and A N Kamal (New York:
Plenum) p 143



\bibitem{Ma}
Ma B -Q 1989 {\it PhD dirssertation} Peking University

 Ma B -Q and Sun J
1990 {\it J.Pkys. G:Nucl. Part. Phys.} {\bf 16} 823; 1990 {\it High
Energy Phys. Nucl. Phys.} (Chinese ed) {\bf 14} 416; 1990 {\it Int.
J. Mod. Phys. A} {\bf 6} 345


\bibitem{Coester}
Coester F 1965 {\it Helv. Phys. Acta} {\bf 38} 7; 1987 {\it
Constraint's Theory and Relativisitic Dynamics} ed G Longhi and L
Lusanna (Singapore: World Scientific) p 159; 1986 {\it The
Three-Body Force in the Three-Nucleon System} ed B L Berman and B F
Gibson (New York:Springer) p 472



\bibitem{Chung}
Chung P L, Coester F, Keister B D and Polyzou W N 1988 {\it Phys.
Rev. C} {\bf 37} 2000\\ Keister B D 1989 {\it Nuclear and Particle
Physics on the Light Cone} ed M B Johnson and L S Kisslinger
(Singapre: World Scientific) p 439


\bibitem{Dirac}
Dirac P A M 1949 {\it \RMP} {\bf 21} 392

\bibitem{Weinberg}
Weinberg S 1966 {\it \PR} {\bf 150} 1313

\bibitem{Wigner}
Wigner E 1939 {\it Ann. Math.} {\bf 40} 149

\bibitem{Isgur}
Isgur N 1980 {\it The New Aspects of Subnuclear Physics} ed A
Zichichi (NewYork: Plenum) p 107 Dziembowski Z 1988 {\it Phys. Rev.
D} {\bf 37} 778

\bibitem{PDG}
Particle Data Group, Yost G P {\it et al} 1988 {\it \PL} {\bf 204B}
1


\bibitem{Close90}
Close F E 1990 {\it \PRL} {\bf 64} 361


\bibitem{Bogolaubov}
Bogoloubov P N  1968 {\it Ann. Inst. H Poincare}  {\bf 8 } 163



\bibitem{Close}
Close F E 1979 {\it An Introduction to Quarks and Partons} (New
York: Academic) p 115, 419


\bibitem{Tegen}
Tegen R 1989 {\it \PRL} {\bf 62} 1724 and references therein



\end{thebibliography}
\end{document}